\documentclass[preprintnumbers,nofootinbib,noshowpacs,eqsecnum,prd,superscriptaddress,letterpaper]{revtex4}

\usepackage[utf8]{inputenc}
\usepackage{graphicx}
\usepackage{amsmath,amssymb}
\usepackage{feynmp}
\usepackage{url}
\usepackage{ulem}
\usepackage{multirow}
\usepackage{hyperref,color}

\newcommand\one{\leavevmode\hbox{\small1\normalsize\kern-.33em1}}

\newcommand{\lag}{\mathcal{L}}

\newcommand{\ope}{\mathcal{O}}
\newcommand{\opp}{\mathcal{P}}

\newcommand{\qqqquad}{\qquad \qquad \qquad}

\def\slashchar#1{\setbox0=\hbox{$#1$}           
   \dimen0=\wd0                                 
   \setbox1=\hbox{/} \dimen1=\wd1               
   \ifdim\dimen0>\dimen1                        
      \rlap{\hbox to \dimen0{\hfil/\hfil}}      
      #1                                        
   \else                                        
      \rlap{\hbox to \dimen1{\hfil$#1$\hfil}}   
      /                                         
   \fi}

\newcommand{\ie}{\textsl{i.e.}\;}

\DeclareMathOperator{\tr}{Tr}

\newcommand{\be}{\begin{eqnarray*}}
\newcommand{\ee}{\end{eqnarray*}}

\newcommand{\bee}{\begin{eqnarray}}
\newcommand{\eee}{\end{eqnarray}}
\newcommand{\beeq}{\begin{equation}}
\newcommand{\eeeq}{\end{equation}}

\begin{document}

\title{The Non-Linear Higgs Legacy of the LHC Run~I}

\author{Tyler Corbett}
\affiliation{C.N. Yang Institute for Theoretical Physics, SUNY, Stony Brook, USA}
\affiliation{ARC Centre of Excellence for Particle Physics at the Terascale,
School of Physics, The University of Melbourne, Victoria, Australia}
\author{Oscar J. P. \'Eboli}
\affiliation{Instituto de Fisica, Universidade de Sao Paulo, Sao Paulo, Brazil}
\author{Dorival Gon\c{c}alves}
\affiliation{Institute for Particle Physics Phenomenology, Department of Physics,
Durham University, UK}
\author{J.~Gonzalez--Fraile}
\affiliation{Institut f\"ur Theoretische Physik, Universit\"at Heidelberg, Germany}
\author{Tilman Plehn}
\affiliation{Institut f\"ur Theoretische Physik, Universit\"at Heidelberg, Germany}
\author{Michael Rauch}
\affiliation{Institute for Theoretical Physics, Karlsruhe Institute of Technology, Germany}

\begin{abstract} 
In the recent paper on \textsl{The Higgs Legacy of the LHC Run~I} we
interpreted the LHC Higgs results in terms of an effective Lagrangian
using the SFitter framework. For the on-shell Higgs analysis of rates and kinematic distributions we relied
on a linear representation based on dimension-6 operators with a simplified fermion sector. In this
addendum we describe how the extension of Higgs couplings modifications in
a linear dimension-6 Lagrangian can be formally understood in terms of
the non-linear effective field theory. It turns out that our previous
results can be translated to the non-linear framework through a simple
operator rotation.
\footnote{This note will be included in the arXiv version of the
  original paper~\cite{Corbett:2015ksa}.}
\end{abstract}

\maketitle
\newpage


In our recent analysis of \textsl{The Higgs Legacy of the LHC
  Run~I}~\cite{Corbett:2015ksa} we have searched for deviations of the
observed Higgs boson from the Standard Model based on two different
parametrizations. First, we studied shifted SM-like Higgs
couplings~\cite{sfitter_orig},
\begin{alignat}{5}
\lag 
= \lag_\text{SM} 
&+ \Delta_W \; g m_W H \; W^\mu W_\mu
+ \Delta_Z \; \frac{g}{2 c_w} m_Z H \; Z^\mu Z_\mu
- \sum_{\tau,b,t} \Delta_f \; 
\frac{m_f}{v} H \left( \bar{f}_R f_L + \text{h.c.} \right) \notag \\
&+  \Delta_g F_G \; \frac{H}{v} \; G_{\mu\nu}G^{\mu\nu}
+  \Delta_\gamma F_A \; \frac{H}{v} \; A_{\mu\nu}A^{\mu\nu} \; .
\tag{1}
\label{eq:lag_delta}
\end{alignat} 
While this $\Delta$-framework\footnote{Our
  $\Delta$-framework~\cite{sfitter_orig} is essentially identical to
  the experimental $\kappa$ framework~\cite{before_sfitter}.} does not
represent a renormalizable field theory outside an effective field
theory framework, it can be linked to a non-linear effective field
theory of the Higgs sector~\cite{non-linear}.\bigskip

In the same analysis~\cite{Corbett:2015ksa}, we expanded this
$\Delta$-framework to a gauge-invariant linear effective Lagrangian, to be
able to include kinematic distributions. Our 9-dimensional operator
basis with the corresponding Wilson coefficients $f_j$
is
\begin{alignat}{9}
  \ope_{GG} &= \phi^\dagger \phi \; G_{\mu\nu}^a G^{a\mu\nu}  
\qqqquad & \ope_{WW} &= \phi^{\dagger} \hat{W}_{\mu \nu} \hat{W}^{\mu \nu} \phi  
\qqqquad & \ope_{BB} &= \phi^{\dagger} \hat{B}_{\mu \nu} \hat{B}^{\mu \nu} \phi 
\notag \\
  \ope_W &= (D_{\mu} \phi)^{\dagger}  \hat{W}^{\mu \nu}  (D_{\nu} \phi) 
& \ope_B &=  (D_{\mu} \phi)^{\dagger}  \hat{B}^{\mu \nu}  (D_{\nu} \phi)
& \ope_{\phi,2} &= \frac{1}{2} \partial^\mu\left ( \phi^\dagger \phi \right)
                            \partial_\mu\left ( \phi^\dagger \phi \right) 
\notag \\
  \ope_t &= \phi^\dagger\phi \; (\overline Q_3 \, \tilde \phi \, t_R)
& \ope_b &= \phi^\dagger\phi \; (\overline Q_3 \, \phi \, b_R) \; ,
& \ope_\tau &= \phi^\dagger\phi \; (\overline L_3 \, \phi \, \tau_R) \; .
\tag{2}
\label{eq:eff}  
\end{alignat}
%
To illustrate how this linear dimension-6 Lagrangian can be phenomenologically viewed as
an expansion of the $\Delta$-framework we spell out the linear
dimension-6 Lagrangian in terms of the physical Higgs field $H$,
\begin{alignat}{5}
\lag 
&= g_{Hgg} \; H G^a_{\mu\nu} G^{a\mu\nu} 
+  g_{H \gamma \gamma} \; H A_{\mu \nu} A^{\mu \nu} 
+ \sum_{f=\tau,b,t} \left( g_f H \bar f_{L} f_{R} + \text{h.c.} \right) 
+ g^{(1)}_{H Z \gamma} \; A_{\mu \nu} Z^{\mu} \partial^{\nu} H \notag \\
&+  g^{(2)}_{H Z \gamma} \; H A_{\mu \nu} Z^{\mu \nu} 
+ g^{(1)}_{H Z Z}  \; Z_{\mu \nu} Z^{\mu} \partial^{\nu} H 
+  g^{(2)}_{H Z Z}  \; H Z_{\mu \nu} Z^{\mu \nu} 
+  g^{(3)}_{H Z Z}  \; H Z_\mu Z^\mu \notag \\
&+ g^{(1)}_{H W W}  \; \left (W^+_{\mu \nu} W^{- \, \mu} \partial^{\nu} H 
                            +\text{h.c.} \right) 
+  g^{(2)}_{H W W}  \; H W^+_{\mu \nu} W^{- \, \mu \nu} 
+  g^{(3)}_{H W W}  \; H W^+_{\mu} W^{- \, \mu} \; .
\tag{3}
\label{eq:linear}
\end{alignat}
%
Of these thirteen $g_{HXX}$ terms modified through the dimension-6
Lagrangian above, seven correspond to the $(1+\Delta_x)$ defined in
Eq.\eqref{eq:lag_delta}, in the custodial limit $\Delta_W = \Delta_Z$,
plus the addition of $\Delta_{Z\gamma}$ if desired~\cite{Corbett:2015ksa}.
Four additional terms $g_{HVV}^{(1,2)}$ add new structures to the
$VVH$ couplings ($V=W,Z$), which can be tested experimentally.\bigskip

The effective Lagrangian based on a non-linear realization of the
electroweak symmetry breaking and including a light scalar $H$ has
been studied in detail in Refs.~\cite{Brivio:2013pma,
  non-linear,non-linear2}. In the non-linear realization the Higgs is
not embedded in a doublet. We can nevertheless link the non-linear and
linear operator sets in terms of canonical dimensions. This connection
is usually established through the ratio of scales $(v/f)^2$, where
$f$ can be related to the scale of strong dynamics.  A detailed
analysis of the non-linear model reveals a double ordering: first,
there is the chiral expansion, and in addition there is the
classification in powers of $(v/f)^2$. Following
Ref.~\cite{non-linear2}, a subset of the non-linear Lagrangian can be
linked to the $\Delta$-framework of Eq.\eqref{eq:lag_delta}, with the
additional assumption $\Delta_W=\Delta_Z$~\cite{sfitter_bsm}. As for
the linear representation, an extended set of non-linear operators
provides a natural extension to include kinematic distributions at the
LHC.\bigskip

For our study, based on~\cite{Brivio:2013pma}, we start with the reduced
bosonic CP-even operator set of order $(v/f)^2$, 
\begin{alignat}{5}
\opp_C &= -\frac{v^2}{4}\tr(\mathbf{V}^\mu \mathbf{V}_\mu) \; \mathcal{F}_C(H) \qqqquad
&\opp_3 &= ig\tr(W_{\mu \nu} [\mathbf{V}^\mu,\mathbf{V}^\nu]) \; \mathcal{F}_3(H) \notag \\
\opp_T &= \frac{v^2}{4} \tr(\mathbf{T}\mathbf{V}_\mu)\tr(\mathbf{T}\mathbf{V}^\mu) \; \mathcal{F}_T(H)
&\opp_4 &= ig' B_{\mu \nu} \tr(\mathbf{T}\mathbf{V}^\mu) \partial^\nu \; \mathcal{F}_4 \notag \\
\opp_H &= \frac{1}{2}(\partial_\mu H)(\partial^\mu H) \; \mathcal{F}_H(H)
&\opp_5 &= ig \tr(W_{\mu \nu}\mathbf{V}^\mu) \partial^\nu \; \mathcal{F}_5(H) \notag \\
\opp_W &=-\frac{g^2}{4} W_{\mu \nu}^a W^{a\mu\nu} \; \mathcal{F}_W(H) 
&\opp_6 &= (\tr(\mathbf{V}_\mu\mathbf{V}^\mu))^2 \; \mathcal{F}_6(H) \notag \\
\opp_G &= -\frac{g_s^2}{4}G_{\mu\nu}^a G^{a\mu\nu} \; \mathcal{F}_G(H)
&\opp_7 &= \tr(\mathbf{V}_\mu\mathbf{V}^\mu) \partial_\nu \partial^\nu \; \mathcal{F}_7(H)   \notag \\
\opp_{B} &=-\frac{g'^2}{4}B_{\mu \nu} B^{\mu \nu} \; \mathcal{F}_B(H) 
&\opp_8 &= \tr(\mathbf{V}_\mu\mathbf{V}_\nu) \; \partial^\mu \mathcal{F}_{8}(H) \; \partial^\nu \mathcal{F}_8'(H)  \notag \\
\opp_{\square H} &=\frac{1}{v^2}(\partial_\mu \partial^\mu H)^2 \; \mathcal{F}_{\square H}(H)
&\opp_9 &= \tr((\mathcal{D}_\mu\mathbf{V}^\mu)^2) \; \mathcal{F}_{9}(H) \notag \\
\opp_1 &= \phantom{\frac{1}{2}}gg' B_{\mu \nu} \tr(\mathbf{T} W^{\mu \nu}) \; \mathcal{F}_1(H)
&\opp_{10} &= \tr(\mathbf{V}_\nu\mathcal{D}_\mu\mathbf{V}^\mu) \partial^\nu\; \mathcal{F}_{10}(H) \notag \\
\opp_2 &= \phantom{\frac{1}{2}}ig'B_{\mu \nu} \tr(\mathbf{T}[\mathbf{V}^\mu,\mathbf{V}^\nu]) \; \mathcal{F}_2(H)  \; ,
\tag{4}
\label{eq:non-linear}
\end{alignat}
where $\mathbf{V}_\mu\equiv
\left(D_\mu U \right) U ^\dagger$ and
$\mathbf{T}\equiv U \sigma_3 U ^\dag$ are the vector and
scalar chiral fields transforming in the adjoint of $SU(2)_L$.  The
dimensionless unitary matrix $U$ contains the Goldstone
modes, $U=e^{i (\sigma \cdot \pi)/v}$, and transforms as a
bi-doublet $ U \rightarrow L\,  U R^\dagger$
under the global $SU(2)_{L,R}$ transformations. The covariant
derivatives are
\begin{align}
D_\mu  U 
&=\partial_\mu U +\frac{i}{2}gW_{\mu}^a\sigma_a U 
                             - \dfrac{ig'}{2} B_\mu  U \sigma_3 \notag \\
\mathcal{D}_\mu \mathbf{V}_\nu 
&=\partial_\mu \mathbf{V}_\nu +i g \left[ W^a_\mu\frac{\sigma_a}{2}, \mathbf{V}_\nu \right] \; ,
\tag{5}
\end{align}
where in the second line $\mathcal{D}_\mu$ is the covariant derivative
for the adjoint representation of
$SU(2)_L$~\cite{Brivio:2013pma}. The model-dependent
functions $\mathcal{F}_i(H)$ introduce the anomalous Higgs couplings.

The number of operators given in Eq.~\eqref{eq:non-linear} can be
further reduced when considering on-shell Higgs
measurements~\cite{Brivio:2013pma}.  First, in analogy to our linear
ansatz we neglect $\opp_1$ and $\opp_T$ because of their tree-level
contribution to the $S$ and $T$ parameters respectively.  Next, operators
containing the combination $\mathcal{D}_\mu\mathbf{V}^\mu$ are
irrelevant for on-shell gauge bosons or when the fermion masses in the process
are neglected. This removes $\opp_9$ and $\opp_{10}$ from our Higgs
analysis.  While $\opp_2$ and $\opp_3$, together with $\opp_4$ and
$\opp_5$, are responsible for one of the interesting de-correlations
that may allow us to distinguish linear from non-linear electroweak
symmetry breaking~\cite{Brivio:2013pma}, they do not affect
three-point Higgs couplings. For the same reason we also omit $\opp_6$
and $\opp_8$.  Finally, for on-shell Higgs amplitudes $\opp_7$ and
$\opp_{\square H}$ lead to coupling shifts~\cite{softening}, \ie
their effect can be accounted for by a re-definition of the remaining
non-linear operator coefficients.\bigskip

Again in analogy to the linear ansatz of Eq.\eqref{eq:linear} we also
add three Yukawa-like non-linear operators of the type
\begin{align}
\opp_t
&= \frac{m_{t}}{\sqrt{2}} \; \overline{Q}_L  U 
   \mathcal{F}_t(H) t_R   +\text{h.c.} \ \ ,
\tag{6}
\label{eq:Opfer}
\end{align}
where a factor $m_{f_i}/v$ has
been introduced with respect to~\cite{Brivio:2013pma}
for a better comparison with~\cite{Corbett:2015ksa}.
With this simplification of the fermion sector we can absorb a
combination of $\opp_C$ and $\opp_H$ in the equations of motion and
arrive at the 9-dimensional non-linear operator set
\begin{align}
\{ \, 
\opp_G, \opp_B, \opp_W, \opp_H, \opp_4, \opp_5, \opp_\tau, \opp_b, \opp_t
\, \} \; ,
\tag{7}
\end{align}
with the appropriate coefficients $c_j$, where the
$(v/f)^2$ factors are implicitly absorbed. Assuming a truncated
polynomial for the Higgs functions of all operators, $\mathcal{F}_i =
1 + 2\tilde{a}_i H/v+...$, and working in the fermion mass
basis we define the non-linear extension to the SM
Lagrangian in analogy to Eq.\eqref{eq:linear}. The combined 
coefficients we denote as $a_j = c_j \, \tilde{a}_j$ for the gauge
operators and $a_j = c_j(2\tilde{a}_j-1)$ for the fermion operators.
This way we can link the non-linear Lagrangian and the linear
Lagrangian relevant for our Higgs analysis,
\begin{align}
\frac{v^2}{2}\frac{f_{BB}}{\Lambda^2} & =  a_B\,,\qquad
& \frac{v^2}{2}\frac{f_{WW}}{\Lambda^2} & =  a_W\,, \qquad
& \frac{v^2}{(4\pi)^2}\frac{f_{GG}}{\Lambda^2} & = a_G\,,  \notag \\
\frac{v^2}{8}\frac{f_{B}}{\Lambda^2} & =  a_4\,, \qquad
& -\frac{v^2}{4}\frac{f_{W}}{\Lambda^2} & =  a_5\,,\qquad
& v^2\frac{f_{\phi,2}}{\Lambda^2} & =  c_H\,,  \notag \\
v^2\frac{f_{t}}{\Lambda^2} & =  a_t\,,\qquad
& v^2\frac{f_{b}}{\Lambda^2} & =  a_b\,, \qquad
& v^2\frac{f_{\tau}}{\Lambda^2} & =  a_{\tau} \, .
\tag{8}
\end{align}
We emphasize that these relations are valid only when we study the
effects of the operators restricted to trilinear Higgs
interactions.

\begin{figure}[t]
  \centering
  \includegraphics[width=0.45\textwidth]{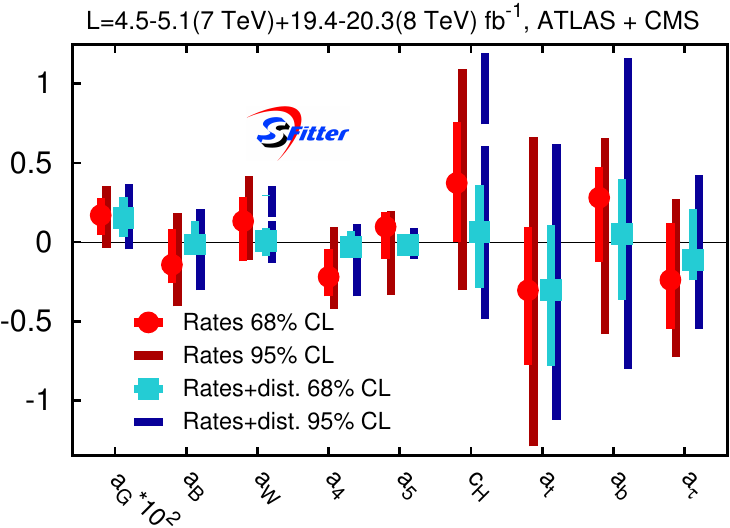}
  \caption{68\% and 95\% CL error bars on the coefficients defined for
    the non-linear operator analysis. The underlying SFitter analysis
    is identical to Ref.\cite{Corbett:2015ksa}.}
\label{fig:non-linear}
\end{figure}

Based on these relations we can express the SFitter Higgs results from
Ref.~\cite{Corbett:2015ksa} in terms of this non-linear subset of
operators. In Figure~\ref{fig:non-linear} we show the 68\% and 95\% CL
allowed regions based on all on-shell Higgs event rates and including
kinematic distributions.\bigskip

Summarizing, in this addendum we have presented the results of the
SFitter Run~I Higgs analysis~\cite{Corbett:2015ksa} in terms of
non-linear effective operators. While the $\Delta$-framework can be
linked to a subset of operators in a non-linear Lagrangian, additional
non-linear operators allows us to also describe kinematic
distributions. This is the same logic as the extension of the
$\Delta$-framework to a linear dimension-6 Lagrangian. If we restrict
our analysis to on-shell Higgs measurements, we find a one-to-one
correspondence between the linear and a non-linear operator set.  The
LHC results in the two approaches can be translated into each other
through a simple operator rotation.

\end{document}